\documentclass[proof]{WileyASNA-v1}

\articletype{Article Type}%
\usepackage{apacite}
\received{7 May 2025}
\revised{10 June 2025}
\accepted{10 June 2025}

\usepackage{graphicx}   
\usepackage{amsmath}    
\usepackage{amssymb}    
\usepackage{longtable}
\usepackage{textgreek}
\usepackage{multicol}
\usepackage{lscape}
\usepackage[T1]{fontenc}
\usepackage{ae,aecompl}
\usepackage[LGR,T1]{fontenc} 
\usepackage{longtable}

\DeclareSymbolFont{greekletters}{LGR}{\familydefault}{m}{n}
\DeclareMathAccent{\dasia}{\mathord}{greekletters}{60}
\DeclareMathAccent{\psili}{\mathord}{greekletters}{62}

\begin{document}

\sloppy

\title{New Arabic records from Cairo on supernovae 1181 and 1006}

\author[1]{J.G. Fischer}
\author[2]{H. Halm}
\author[3]{R. Neuh\"auser (*)}
\author[4]{D.L. Neuh\"auser}

\authormark{Fischer \textsc{et al.} Supernovae 1181 and 1006}

\address[1]{Institut f\"ur Arabistik und Islamwissenschaft, Universit\"at M\"unster, Schlaunstrasse 2, 48143 M\"unster, Germany}
\address[2]{Asien-Orient-Institut, Universit\"at T\"ubingen, Wilhelmstrasse 113, 72070 T\"ubingen, Germany}
\address[3]{\orgdiv{Astrophysikalisches Institut}, \orgname{Universit\"at Jena}, \orgaddress{\state{Schillerg\"asschen 2-3, 07745 Jena}, \country{Germany}}}
\address[4]{Independent scholar, Venezia, Veneto, Italy}

\corres{*Corresponding author name: Ralph Neuh\"auser. \email{ralph.neuhaeuser@uni-jena.de}}
\corres{*Ralph Neuh\"auser. \email{ralph.neuhaeuser@uni-jena.de}}

\abstract{The remnant of the historical supernova SN 1181 is under 
discussion: While the previously suggested G130.7+3.1 (3C58) appears too old 
(3000-5000 yr), the unusual star IRAS 00500+6713 with a surrounding 
nebula (Pa-30) has an expansion age 
not inconsistent with a SN Iax explosion in AD 1181 under the assumption that neither acceleration nor deceleration occurred.
Previously, only reports from China and Japan were known, pointing to 
an event near the northern circumpolar region. 
Any further reports from other cultures can therefore be highly relevant. 
We present here an Arabic poem in praise of Saladin by the 
contemporaneous author Ibn San\={a}' al-Mulk (Cairo, Egypt). 
We re-date its composition to between Dec 1181 and May 1182. 
It contains a new bright star, which can be identified as SN 1181. 
The poem also provides new and independent information on 
the object type (called `najm' for `star'), location on sky 
(in or near the Arabic constellation al-Kaff al-Kha\d{d}īb, lit. the henna-dyed hand 
(five bright stars in Cassiopeia), and brightness (brighter than $\alpha$ Cas, 2.25 mag). 
In addition, we present another Arabic text on SN 1006, also from Cairo, by the historian 
al-Maqrīzī, probably based on the contemporaneous al-Musabbi\d{h}ī.}


\maketitle


\section{Introduction: Historical Supernovae}

We summarize the importance of historical supernovae (Sect. 1.1) 
and the relevant facts known for the new stars (supernovae) of AD 1181 (Sect. 1.2) and AD 1006 (Sect. 1.3), 
for both of which we present here newly found Arabic observations.

\subsection{Relevance of historical supernovae for astrophysics}

There are two main supernova (SN) types, namely core-collapse supernovae (cc-SNe) of massive stars 
after the end of their fusion phase, often forming neutron stars, and thermonuclear SNe (type Ia) 
of White Dwarfs in single- (White Dwarf plus normal star) or double-degenerate (two White Dwarfs) binaries. 
SNe can be observed even by the unaided eye if within a few kpc. Indeed, several such explosions 
have been recorded before the invention of the telescope, e.g. in AD 1604, 1572, and 1006 (all Ia). Examples for 
remnants of Galactic cc-SNe are the Crab pulsar and the SN remnant (SNR) Cas A with its central compact object -- 
however, the explosion around AD 1680 producing Cas A probably remained unnoticed. 
See Stephenson \& Green (2002, henceforth SG02) for a review on historical SNe.

While all known historical SNe have been observed by court astronomers in China, some were also detected in Japan, Korea, Europe, and Arabia (SG02) -- see also Rada \& R. Neuh\"auser 
(2015) and R. Neuh\"auser et al. (2016, 2017a) for newly found observations of SNe 1006, 1572, and 1604 from Yemen as well as R. Neuh\"auser et al. (2017b) for Ibn Sīn\=a's observation of SN 1006 from Persia. SNe 1572 and 1604 were observed extensively in Europe, e.g. by Tycho Brahe in 1572/73 as well as by David Fabricius and Johannes Kepler in 1604/05. Stationarity relative to the fixed stars, star-like scintillation, and non-detection of a parallax (nor orbital motion in the solar system) helped to establish in the Early Modern Period that these transient celestial objects are located not only outside the Earth atmosphere (supra-lunar), but even beyond the solar system 
(e.g. Weichenhahn 2004 for SN 1572 and Boner 2020 for SN 1604).

Stephenson \& Green (2005) listed the following criteria for historical SNe:  \\
(i) `long duration ... (preferably more than three months)',  \\
(ii) `fixed location' relative to stars,  \\
(iii) `low Galactic latitude (usually less than $10^{\circ}$)',  \\
(iv) `no evidence of significant angular extent (... no indication of a tail)',  \\
(v) `unusual brilliance (e.g. daylight visibility or brightness comparable to one of the brighter planets)', and  \\
(vi) `several independent records'.  \\
Similar criteria apply to novae, but with lower brightness and, hence, shorter duration 
(unless very nearby), and wider distribution around the Galactic plane.

Historical SNe are important for astrophysics: Their exact age is known (i.e. the time since the explosion, which is usually dated in the historical transmission to within a few days) with exceedingly better precision and accuracy than any astrophysically determined age of neutron stars or SNRs; it is similar also for historical novae and their shells. Sometimes, historical observations can provide a light curve and/or even a color evolution (both can fix the SN sub-type, i.e. which kind of star or stars exploded), or at least the peak brightness, which can (together with the SN type) determine the distance.
And given that they were observed by the unaided eye, they were nearby, so that detailed telescopic follow-up
observations of their remnants are now possible.

Historical observing reports on `new stars' (Latin: novae) or `guest stars' (Chinese: ke xing)
can be found in chronicles and annals. 
The term `cometes' had a more general meaning in pre-telescopic times,
not only non-stationary ones with tails (today: comet),
but also transient stationary non-extended objects (today: nova or SN).
E.g., SN 1006 was called a `comet' (e.g. SG02, p. 169). 
Likewise, SN 1572 was called `comet' in some European publications 
(e.g. by its first European observer, Jerome Munoz, SG02, p. 90) as well as the variable star Mira
by Fabricius and Brahe (R. Neuh\"auser \& D.L. Neuh\"auser et al. 2024).  

Similar, the Arabic term `nayzak’ (for `short spear’) 
can refer to either a nova/supernova or a comet (e.g. R. Neuh\"auser et al. 2016). 

Goldstein (1965) was the first to find an Arabic record on a SN: 
He noticed that what was previously considered the comet of the year AD 1006 was actually a SN, 
here reported by $^{c}$Alī ibn Ri\d{d}w\=an from Cairo, Egypt, in Arabic language using terms 
like `nayzak’ and `athar' for `spectacle' (ca. $-7.5$ mag at peak);
other Arabic authors used `kawkab' (today: star or planet) or 'najm' (today: star) for SN 1006 (Goldstein 1965). 
Later, several more reports on SN 1006 from Arabia and Persia were found
(Rada \& R. Neuh\"auser 2015, R. Neuh\"auser et al. 2017a,b).
They also inform about position, duration of visibility, and strong brightness.

\subsection{The transient of AD 1181 (SN 1181)}

The transient of AD 1181/82 is relatively well documented in sources from China and Japan
(Hsi 1957, Xu et al. 2000, SG02).
In south China, this `guest star' is reported
from 1181 Aug 6 until 1182 Feb 6 (185 nights), 
in northern China from 1181 Aug 11 on for 156 days,
and in Japan from Aug 7 until at least Sep 26 (SG02).

Since several of the above mentioned SN criteria are fulfilled (long duration, low Galactic latitude, 
and several independent records), the transient was suggested since long as a candidate of a SN or slow nova (e.g. Hsi 1957, Stephenson 1971). 
Henceforth, we consider the transient of AD 1181/82 as a stellar object -- it is still mostly considered
as supernova (SN 1181), e.g. SG02 or Schaefer (2023), but a slow nova is also not impossible.

Given that the new star of AD 1181/82 is recorded by several sources from China and Japan, 
that it was seen in total for about half a year, and its possible comparison with Saturn in brightness, 
SG02 tentatively estimated its peak brightness to be around 0 mag.

Stephenson (1971) suggested the faint SNR G130.7+3.1 (3C58) for SN 1181, 
whose emission is dominated by a Pulsar Wind Nebula related to pulsar PSRJ0205+6449;
Fesen et al. (2008) estimated the age of this SNR to be 3000-5000 yr. 
The spin-down age (i.e. upper age limit) of the pulsar is 5370 yr (Livingstone et al. 2009), 
which is clearly better consistent with the SNR age than the time span since AD 1181. 
Hence, remnant and the type of the transient of AD 1181/82 would then be unknown.

Recently, Oskinova et al. (2020) and Ritter et al. (2021) suggested the star IRAS 00500+6713 
(with fast winds) and its surrounding nebula called Pa-30 (expansion age ca. $1000 \pm 250$ yr), 
located in the general direction considered for the guest star of AD 1181, 
as remnant from a double-degenerate SN Iax explosion\footnote{Note that all extra-galactic SN Iax are suggested to occur 
as pure-deflagration explosion triggered 
by helium accretion onto one White Dwarf (single degenerate), see e.g. the review in Jha (2017), 
while the star Pa-30 as suggested remnant of SN 1181 would have resulted from a 
merger of a CO and a ONe White Dwarf (double degenerate), see Schaefer (2023).
However, the extra-galactic SNe Iax and the suggested scenario for Pa-30 all show low explosion energies and low expansion velocities.} in AD 1181. 
Fesen et al. (2023) detected radially aligned filaments in Pa-30, 
which would be consistent with an explosion around AD 1181,
if there was neither deceleration nor acceleration;
both would be expected, though, due to the interstellar medium and the hot star wind.
Furthermore, Schaefer (2023) re-determined the position using the translations from Chinese in SG02
and then concluded that SN 1181 is connected to IRAS 00500+6713 / Pa-30 
and that it was a (double-degenerate) SN type Iax from a merger of a CO and a ONe White Dwarf. 
However, Lykou et al. (2023) found the Ne to O abundance ratio in spectra of Parker's star to be too low for an ONe White Dwarf.
Searching for the remnant star on photographic plates, Schaefer (2023) found that its drop 
in brightness varried between 0.03 mag/decade from 1950-2012 and 0.21-0.22 mag/decade from 1889-1950 and 2012-2022, respectively,
similar in Lykou et al. (2023), but only 0.01 mag/yr in WISE W1 and W2 bands from 2011 to 2020 (Lykou et al. 2023),
so that a linear long-term extrapolation back to AD 1181 might be problematic.
The immediate surroundings of IRAS 00500+6713 / Pa-30 were more recently also observed in X-rays by Ko et al. (2024),
who need an extremely fine-tuned start of the Pa-30 wind in the 1990ies for their model,
and with optical integral field spectroscopy by Cunningham et al. (2024), 
who assumed that there was neither acceleration nor deceleration in the remnant to link Pa-30 to SN 1181.

Stephenson (1971) and Schaefer (2023) both interpreted the same records from China,
but suggested positions (3C58 and Pa-30), which are ca. $8^{\circ}$ apart -- this may indicate that
the records are imprecise and/or that the search area is relatively large in this case.
Given that object type (nova, SN) and remnant may still be uncertain 
and that definite position and brightness are still not sufficiently precisely determined
from the historical records, any new information on this transient
in particular on its position on sky, brightness/color, or object type (comet or fixed star),
would be highly welcome and relevant.

\subsection{The supernova of AD 1006}

$^{c}$Alī ibn Ri\d{d}w\=an from Cairo (Egypt) reported the new star for AD 1006 with quite some details: 
(a) duration May to July 1006, 
(b) location in the 15th degree of Scorpion, i.e. an ecliptic longitude range, 
(c) a brightness of `2.5-3 times as large as Venus' and `more than a quarter of that of moonlight', and 
(d) stationarity (`it remained where it was ...'), see SG02, p. 163. 
Several more Arabic records on SN 1006 are consistent with $^{c}$Alī ibn Ri\d{d}w\=an's, 
including appearance around early May 1006, but report a total visibility of up to 6 months (SG02). 
In addition, Ibn Sīn\=a seems to report a color evolution (R. Neuh\"auser et al. 2017b). 
Two records from Yemen date the first appearance clearly to mid-Rajab of AH 396, i.e. 1006 Apr $17 \pm 2$ (Rada \& R. Neuh\"auser 2015). 
In addition, a monk from St. Gallen reports to have seen the object `for 3 months' at the `limits of the south'; 
if he would have seen it (i) for at least one minute per night after sunset, 
(ii) at least one degree above the mountain range, and (iii) for at least 2.5 months (rounded to `3 months'), 
then he would have started to see it around AD 1006 Apr 25 or earlier 
(R. Neuh\"auser et al. 2017a). The guest star of AD 1006 is also reported in various Chinese and Japanese sources 
since April or May 1006 for up to 2 years (Xu et al. 2000).

From the ecliptic latitude range given by $^{c}$Alī ibn Ri\d{d}w\=an, the right ascension range reported from China, and the lower declination limit from St. Gallen, the SNR could be identified as G327.6+14.6 (see SG02). Since no neutron star or otherwise compact object is related to the SNR of this very bright and, hence, very nearby SN 1006, it was classified as type Ia (thermonuclear). 
Furthermore, a distance of $2.18 \pm 0.08$ kpc was derived from the proper motion of the ejecta and the shock velocity of filaments (Winkler et al. 2003). At its small distance, any former companion or donor star would have been found, so that from the lack of such a star, it was concluded that SN 1006 was a double-degenerate SN Ia, i.e. a merger of two White Dwarfs. As SN Ia (peak absolute brightness $-19$ mag) at the given distance, at peak, it should have had an apparent brightness of ca. $-7.5$ mag, fully consistent with $^{c}$Alī ibn Ri\d{d}w\=an's record. 

\bigskip

The setup of our work is as follows: In Sect. 2, we will present a newly found observation of the new star of AD 1181/82, 
namely in a poem by Ibn San\=a' al-Mulk from Egypt, completed between Dec 1181 and May
1182 in order to praise Saladin. This is the first observation of SN 1181 written in Arabic language (from Cairo). 
We will first introduce the historical background on Saladin in Sect. 2.1, then the author and his poem in Sect. 2.2, 
also the dating of the poem (2.3), then we bring the full text line by line in Arabic and our English translation (2.4), 
and then comment and discuss the astronomical content in Sects. 2.5 \& 2.6.
In Sect. 3, we also present more briefly a new Arabic observation of SN 1006. We summarize our findings in Sect. 4.

\section{New Arabic record on AD 1181 from Egypt}

Let us now consider the historical background, the author and his work, the dating of the poem, and the text itself.

\subsection{Historical background: Saladin}

Around AD 1181, the Middle East was undergoing a political transformation that would leave its mark on the region for centuries to come. 
One person in particular stood at the centre of this transformation: al-Malik al-N\=a\d{s}ir \d{S}al\=a\d{h} al-Dīn Y\=usuf ibn Ayy\=ub, 
better known in the West as Saladin (for an extensive biography see Lyons \& Jackson 1982, on which this paragraph is based). 
Born in AD 1138 in Tikrit (modern-day Iraq) to a family of Kurdish military officers originally from Dvin (modern-day Armenia), 
Saladin had grown up in Damascus (modern-day Syria). 
There his father and uncle held important positions in the service of N\=ur al-Dīn, 
who ruled over those parts of the Levant that were not in the hands of the Crusaders. 
When N\=ur al-Dīn set his sights on bringing Egypt under his control as well, he sent Saladin's uncle Shīrk\=uh on three expeditions there. 
The third expedition in AD 1168-1169 was finally crowned by success, and after Shīrk\=uh's premature death two months later, Saladin, 
who had accompanied his 
uncle to Egypt, became N\=ur al-Dīn's deputy in the newly conquered country. 
Saladin proved very adept at expanding and consolidating his power; when N\=ur al-Dīn died in AD 1174, leaving an underage son as his heir, 
Saladin occupied Damascus and went on to spend the following decade trying to reassemble N\=ur al-Dīn's former domains under his own rule, 
a process which would culminate in the capture of Aleppo (Syria) in AD 1183. 
The crowning success of his career, however, would be the crushing defeat of the Crusader armies at Hattin and subsequent conquest of Jerusalem in AD 1187. 
At his death in AD 1193, Saladin left a sizable empire encompassing much of the Middle East, which would be ruled by members of his family,
the Ayyubids, until the mid-13th century, when it was taken over by the Mamluks. 
The latter would continue to build on the foundations laid by N\=ur al-Dīn and Saladin 
until their Empire would succumb to the conquering Ottoman armies in AD 1516/1517.

Saladin's life and career coincided with an unprecedented flowering of literature, especially poetry, in Egypt and the Levant. 
Saladin and his close associates alone patronized dozens of poets, and the comparatively high rates of literacy among the urban male population,
combined with the formation of a scholarly infrastructure of madrasahs (colleges) 
and public libraries meant that poets and poetry could find appreciative audiences 
even outside of court circles. The poets of this period set new standards that would shape the subsequent development 
of Arabic poetry, and some of them attained the status 
of classics whose work was widely read and imitated until the 20th century. 
One of these was Ibn San\=a' al-Mulk, the author of the text that concerns us here. 
Before we turn to this text, however, a brief summary of the author's biography, based on Skoog (2011), is in order (another 
useful overview is provided by $^{c}$Abd al-\d{H}aqq in the introduction to his edition of the dīw\=an).

\subsection{Ibn San\=a' al-Mulk: Author and text}

Al-Q\=a\d{d}ī al-Sa$^{c}$īd $^{c}$Izz al-Dīn Ab\=u l-Q\=asim Hibat All\=ah ibn Ja$^{c}$far ibn San\=a' al-Mulk (or just Ibn San\=a' al-Mulk for short) 
was born in Cairo (Egypt) around AD 1155. He came from a respected and well-connected family and received a rounded education that allowed him to pursue a successful career as a secretary 
in the administration of Saladin's blossoming empire. Another important factor in his career was his close relationship with his mentor and protector 
al-Q\=a\d{d}ī al-F\=a\d{d}il, Saladin's right-hand man and the most powerful individual in his realm after the sultan himself. 
Al-Q\=a\d{d}ī al-F\=a\d{d}il also served as intermediary between Saladin and Ibn San\=a' al-Mulk, 
presenting most if not all of the poems the poet dedicated to the sultan to the latter. 
Although Ibn San\=a' al-Mulk's success gained him the envy of many colleagues who tried to sully his name, 
he managed to maintain his position even after Saladin's death and under several subsequent rulers. 
In AD 1209 the poet was offered control over the Military Office (Dīw\=an al-Jaysh), but in light of his advanced age he declined. 
He died two years later in AD 1211. See Skoog (2011) for more details.

One of the genres in which Ibn San\=a' al-Mulk excelled was panegyric or praise poetry, which he dedicated primarily to 
al-Q\=a\d{d}ī al-F\=a\d{d}il and to powerful members of the Ayyubid dynasty. One of these poems contains the mention of
a new {\it `najm'} (star) in the constellation al-Kaff al-Kha\d{d}īb (`Dyed Hand'), made up of five bright stars of Cassiopeia.
It is written in the form of a qa\d{s}īdah, a long, poly-thematic 
poem that maintains a uniform metre and rhyme throughout; in this case, 56 verses in the metre \d{t}awīl rhyming in -m\=a. 
The first 16 verses of the text, which can be found transcribed and translated below, strive to establish an astrological link between 
Saladin's earthly rule and the celestial order, taking the transient as their starting point. The use of such astrological notions was quite 
common in political panegyrics in the pre-modern Islamic world: since the heavens were widely thought to predict and/or cause events on earth,
poets emphasized the ruler's power by portraying him as either enjoying the favor of the stars or even ruling over them as if they were his subjects.
The rest of the poem for the most part praises Saladin for possessing the three essential virtues of the pre-modern Islamic ruler:
munificence, justice, and military prowess. In terms of content, this poem can therefore be said to adhere quite closely to the conventions of Arabic 
praise poetry. What sets it apart is its extraordinary literary quality, which, however, shall not detain us here.

We have already mentioned that Ibn San\=a' al-Mulk's poetry continued to enjoy great popularity in the centuries after his death. 
His dīw\=an (collected poetic works) survives in a great number of manuscripts housed in libraries all over the world, 
and two modern print editions are available, one published in India in 1958, prepared by Mu\d{h}ammad $^{c}$Abd al-\d{H}aqq, 
the other in Egypt in 1969, prepared by Mu\d{h}ammad Ibr\=ahīm Na\d{s}r and \d{H}usayn Mu\d{h}ammad Na\d{s}\d{s}\=ar. 
Unfortunately, neither of these fulfills contemporary standards of a critical scholarly edition, which is why the text presented below is
derived from a new critical edition of the whole poem based on all available manuscripts (to be published in Fischer, in preparation). 
Despite the shortcomings of the published editions, Ibn San\=a' al-Mulk's poetry, including the poem that concerns us here, 
can be considered to be relatively accessible and well-known for an Arabic poet of the period. 
Yet nobody seems to have noticed that it contains a mention of the transient of AD 1181/82.
This is partly due to the fact that the poem has so far been dated incorrectly.

\subsection{Dating the poem}

The manuscript sources for our poem do not assign any dates to its composition. $^{c}$Abd al-\d{H}aqq, the first editor of the dīw\=an, 
thought the poem referred to an event in AD 1186, in which, according to many astrologers, a conjunction of 
all planets around the Sun, i.e. invisible, 
would lead to the end of the world (Ibn San\=a' al-Mulk 1958, p. 650 and pp. 24f). 
For this event see Lyons \& Jackson (1982, p. 246), Weltecke (2003), and M\"ohring (1997, pp. 192f).
Both modern editors (1958 and 1969) may have considered the theory of `Great Conjunctions' by Ab\=u Ma$^{c}$shar (9th century AD), 
according to which conjunctions of planets could create new celestial objects like comets. 
Only Bauer (2011, p. 304) expressed some doubts about the validity of the previous dating to AD 1186, but offered no alternatives.

The confusion about the exact astronomical event that the poem might refer to is compounded by the information provided in the manuscript sources: 
in one manuscript and both editions of the dīw\=an, the introductory remark in prose that serves to contextualize the poem speaks of a “star with a tail” (kawkab lah\=u dhu'\=abah),
a common term for comets (e.g. Rashed 2017, p. 742f). However, this choice of words should not be given too much importance, 
since it is unlikely to have originated with the 
author of the poem or even the observer(s) of the transient: the dīw\=an of Ibn San\=a' al-Mulk was compiled (most likely after his death) 
in at least two more or less independent recensions, and the authorship of the introductions is unknown; the fact that their content and extension can vary
considerably between manuscripts indicates that at least some of them are the work of the compilers or even later copyists;\footnote{The textual 
history of Ibn San\=a' al-Mulk's dīw\=an will be elucidated in Fischer (in preparation).} 
regarding the astronomical part of the poem we discuss here, there are no significant differences between the two recensions. 
In any case, the poem itself never uses the expression {\it kawkab lah\=u dhu'\=abah}, but consistently speaks of a {\it najm} (star). 

Our new dating rests on two passages in the poem: \\
(a) In the last three verses of the poem, Saladin's brother (and eventual successor) 
al-Malik al-$^{c}$\=Adil Sayf al-Dīn Ab\=u Bakr, 
known to medieval Latin chroniclers as Saphadin, is praised alongside Saladin. 
The most likely explanation for this is that the poem was meant to be recited on an occasion when both men were present. 
When Saladin had left Egypt after N\=ur al-Dīn's death in AD 1174, 
al-$^{c}$\=Adil 
had stayed behind and remained in Egypt until after the conquest of Aleppo in 1183 
(Lyons \& Jackson 1982, p. 208 and passim). 
The poem would therefore have been written during one of 
Saladin's stays in Egypt during this period, 
something which is further confirmed by the prominent mention of Egypt in v. 49-52 of the poem. 
Saladin last stayed in Egypt between 1181 Jan 2 and 1182 May 11 
(Lyons \& Jackson 1982, pp. 153-165). \\
(b) In v. 31-32, Saladin's troops are said to protect the holy sites of Mecca from 
destruction.\footnote{v. 31-32: `His soldiers defend the Inviolable House [Kaabah in Mecca] ...
al-\d{H}a\d{t}īm [specific part of the Meccan sanctuary] would have become crushed'.}
This passage does not belong to the standard repertoire of Arabic 
panegyric in this period.
The poem most likely alludes to one of two failed Crusader 
attacks on the Hejaz region, both of them organized by Reynald of Ch\^{a}tillon,
the Lord of Kerak in modern-day Jordan. The poem would thus have been written after one of these two attacks. 
For accounts and analyses of these raids, see Hamilton (1978, pp. 102-104), Lyons \& Jackson (1982, pp. 157f, 160-162 and 185-188), 
Hamilton (2000, pp. 170f and 179-185), and Mallett (2008). The first of these attacks
occurred in 1181 December and followed the inland caravan routes to Tabuk, 
an important stop on the way the pilgrimage caravans took to Medina.
Its purpose was to prevent Saladin's troops in Syria from attacking Aleppo, which was leaderless in 
the period 1181 December 2 to 29 (Hamilton 2000, pp. 170/171), 
by diverting their attention towards the south.
The only pre-modern chronicler to date this raid with some degree of precision, al-Maqrīzī (AD 1364-1442, see also Sect. 3), places it (or at least its beginning)
in his entry dealing with the month of Rajab AH 577 (AD 1181 Nov 10 to Dec 9), see Broadhurst \& al-Maqrīzī (1980, p. 66). 
While Mecca and Medina themselves were not attacked directly by the Franks, who in any case had probably never intended to go that far south, 
official proclamations and letters emanating from Saladin's chancery presented the successful repulsion of this raid by Ayyubid troops 
as having prevented the destruction of Islam's holy sites. This official discourse is reflected in v. 31-32 of the poem.
There may have been more hostilities in early 1182 AD.\footnote{Hamilton (1978): `Probably in the early spring of 1182 he [Reynald] set out to attack 
the oasis of Teima, 250 miles south of Kerak, on the pilgrimage route to Medina. The governor of Damascus intervened and forced him to retreat,
but he was nevertheless able to capture a rich pilgrim caravan' -- the pilgrimage was in the month of Dh\=u al-\d{H}ijjah (AD 1182 Apr 7 to May 6).
To date the first raid to the early spring of 1182 is at odds with Hamilton's own explanation of its goals, as by that time the danger to Aleppo had passed. 
Al-Maqrīzī actually mentions this raid twice, once in the entry on Rajab 577 mentioned above, and once at the beginning of the chapter on the year AH 577 
(Broadhurst \& al-Maqrīzī 1980, p. 63). This is due to the fact that his chronicle combines information from several different sources 
that date these events with varying degrees of precision.}
Either way, the earliest possible date for the first attack -- probably alluded to in the poem -- remains December 1181.
The second attack, this time by ship, happened in the Red Sea in the winter of AD 1182/83, after Saladin had already left Egypt (see a), i.e. too late.

Taken together, these two pieces of information indicate that Ibn San\=a' al-Mulk wrote
and/or completed this poem in or soon after 1181 December and recited it before 1182 May 11.

\subsection{The poem on Saladin with the transient of AD 1181/82}

Below we provide a transcription and commented translation of the relevant passage of the poem. 
The text is based on a new critical edition of the entire poem to be published in Fischer (in preparation). 
Aside from the much shorter prose introduction, however, this new edition does not present major differences from the texts of the two 
existing print editions (Ibn San\=a' al-Mulk 1958, pp. 650-657; Ibn San\=a' al-Mulk 1969, pp. 270-273). 
See Fig. 1 for the relevant folio 165r of one of the manuscripts.

Wa-q\=ala yamda\d{h}u l-Malika l-N\=a\d{s}ira: \\
He said in praise of al-Malik al-N\=a\d{s}ir [Saladin]:\footnote{In the other recension of Ibn San\=a' al-Mulk's dīw\=an, this prose introduction goes as follows: 
Wa-k\=ana qad \d{h}adatha fī l-sam\=a'i fī l-Kaffi l-Kha\d{d}ībi kawkabun lah\=u dhu'\=abatuh\=u wa-lam tajri l-$^{c}$\=adatu bi-\d{z}uh\=uri mithlihī fa-q\=ala 
yamda\d{h}u l-Malika l-N\=a\d{s}ira wa-yadhkuru l-kawkaba lladhī \d{z}ahara (In the sky in the Dyed Hand [Cassiopeia], 
a star with a tail had appeared, 
the like of which had not been seen before, so he said in praise of al-Malik al-N\=a\d{s}ir, mentioning the star that had appeared).} 

\noindent 1. Ar\=a kulla shay'in fī l-basī\d{t}ati qad nam\=a / bi-$^{c}$adlika \d{h}att\=a namat anjumu l-sam\=a. \\
I see how everything on the surface of the Earth has increased in number thanks to your justice; now even the stars [anjum] in the sky have increased in number. 

\noindent 2. ta\d{h}allat bi-najmin l\=a bali btasamat bihī / wa-man sarrah\=u shay'un yasurru tabassam\=a. \\
\noindent [The sky] 
adorned itself with a star [najm]; nay, it smiled through it, because whoever is delighted by a delightful thing smiles. 

\noindent 3. wa-m\=a bari\d{h}a l-Kaffu l-Kha\d{d}ību mu$^{c}$a\d{t}\d{t}alan / fa-lamma ta\d{h}all\=a l-dahru minka takhattam\=a. \\
The Dyed Hand [Cassiopeia] used to be unadorned, but when time became adorned because of you, it put on a (signet) ring.

\noindent 4. fa-l\=a yaftakhir kaffu l-sam\=a'i bi-najmihī / fa-kam a\d{t}la$^{c}$at af$^{c}$\=aluka l-ghurru anjum\=a. \\
Let not the sky’s hand boast of its star, since so many of your noble deeds shine bright like stars. 

\noindent 5. nuj\=umuka m\=a a$^{c}$yat $^{c}$al\=a r\=a\d{s}idin lah\=a / wa-dh\=a l-najmu a$^{c}$y\=a r\=a\d{s}idan wa-munajjim\=a. \\
Your stars have never perplexed any astronomer, while this star has perplexed astronomers and astrologers alike. 

\noindent 6. takh\=alafati l-aqw\=alu fīhi wa-jamjamat / wa-lam nara qawlan fī ma$^{c}$\=alīka jamjam\=a. \\
They are in disagreement over it and stammering; but never have we seen anyone stammer when it came to your high station. 

\noindent 7. nar\=aka naqalta l-rum\d{h}a fī l-ufqi r\=aki\d{d}an / fa-abqayta zujjan thumma alqayta lahdham\=a. \\
We can see how you buried a spear on the horizon, leaving its butt behind and throwing its sharp head.

\noindent 8. wa-dh\=a ghala\d{t}un min fikratī idh takhayyalat / wa-dh\=a kha\d{t}a'un min kh\=a\d{t}irī idh tawahham\=a. \\
But this is just an error due to my fantasy, and a mistake due to my illusions. 

\noindent 9. ab\=uka huwa l-najmu lladhī min ma\d{h}allihī / ta\d{t}alla$^{c}$a musht\=aqan ilayka musallim\=a.  \\
\noindent [In reality] 
the star is your father, who left his abode out of longing for you and in order to greet you. 

\noindent 10. nu\d{s}irta bi-afl\=aki n-nuj\=umi fa-shuhbuh\=a / khamīsun bihī turdī l-khamīsa l-$^{c}$aramram\=a. \\
The stars' spheres come to your aid and their meteors are an army with which you destroy mighty armies. 

\noindent 11. fa-kam ashra$^{c}$a l-rum\d{h}a l-Sim\=aku mu\d{t}\=a$^{c}$inan / $^{c}$aduwwaka \d{h}att\=a k\=ada an yata\d{h}a\d{t}\d{t}am\=a. \\
How often did Arcturus brandish its spear to strike your enemy until it was almost shattered to pieces! 

\noindent 12. wa-m\=a man ghad\=a fī \d{s}af\d{h}ati l-ar\d{d}i \d{h}\=akiman / ka-man \d{z}alla fī shuhbi l-sam\=a'i mu\d{h}akkam\=a. \\
Those who rule on the surface of the earth cannot be compared to one destined to rule the lights of the sky. 

\noindent 13. raqīta il\=a an lam tajid laka murtaqan / wa-aqdamta \d{h}att\=a lam tajid mutaqaddam\=a. \\
You rose until it was impossible to rise any further; you progressed until it was impossible to progress any further. 

\noindent 14. fa-m\=a yubrimu l-miqd\=aru m\=a kunta n\=aqi\d{d}an / wa-m\=a yanqu\d{d}u l-miqd\=aru m\=a kunta mubrim\=a. \\
Fate does not allow what you refuse to happen, and neither does it refuse what you allow. 

\noindent 15. fidan li-bni Ayy\=uba l-nuj\=umu fa-innahum / lah\=u khadamun yafd\=una minhu l-mukhaddam\=a, \\
May the stars sacrifice themselves for the son of Ayy\=ub [Saladin], for they are his servants and thereby sacrifice themselves for the master, 

\noindent 16. wa-m\=a z\=ala a$^{c}$l\=a bi-l-mak\=anati minhum\=u / wa-m\=a z\=ala minhum bi-l-hid\=ayati a$^{c}$lam\=a. \\
who forever stands above them in eminence and forever is a better guide than they are.

\noindent 17. fa-l\=a taqrin\=uh\=u bi-l-mul\=uki fa-innah\=u / ajalluhum\=u ar\d{d}an wa-a$^{c}$l\=ahum\=u sam\=a. \\
So do not put him on the same level as other princes, for he is the greatest of them on earth and the most elevated of them in the sky.

\begin{figure}
\centering
\includegraphics[angle=0,width=\columnwidth]{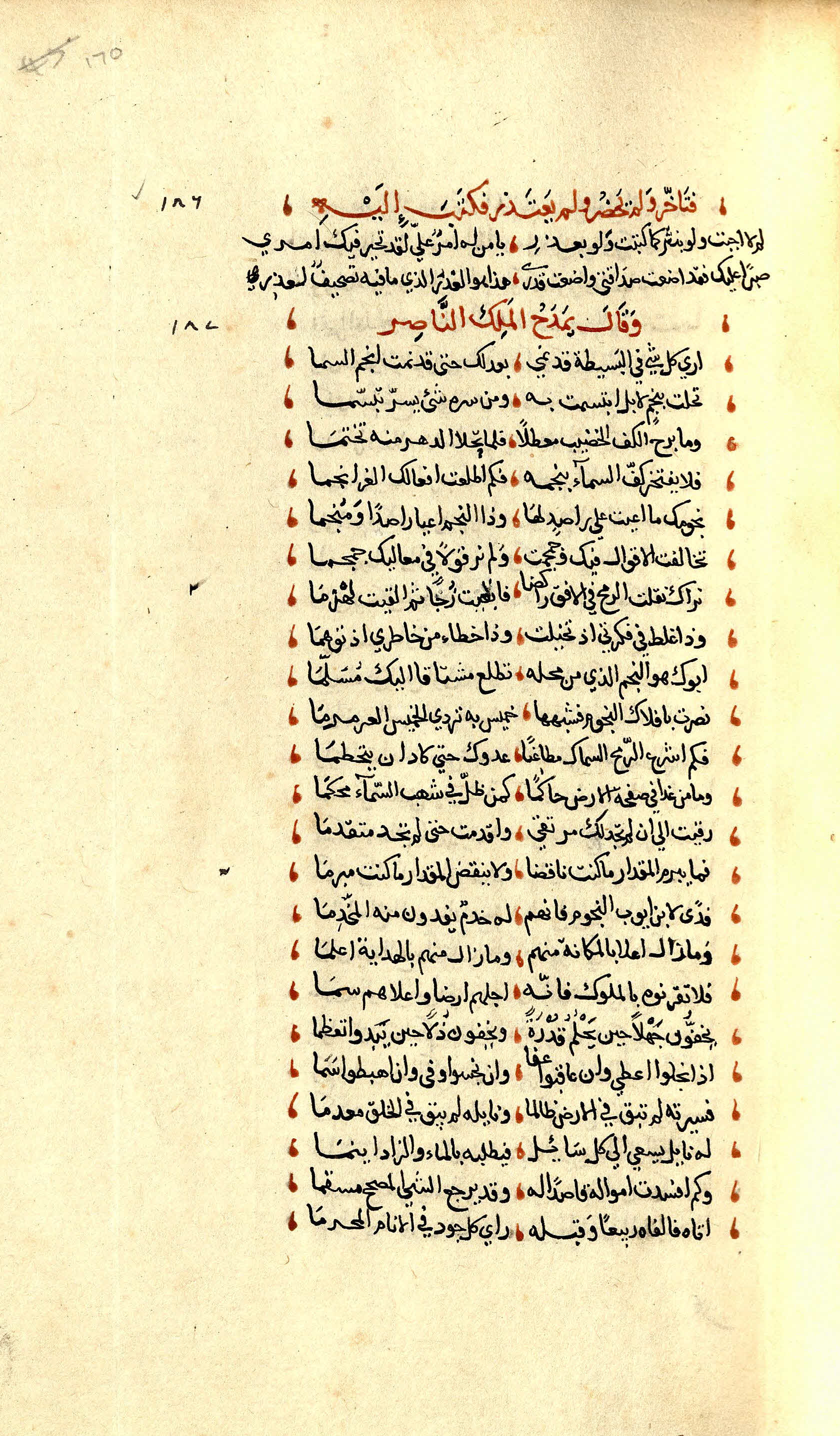}
\caption{{\bf Arabic text SN 1181:} Scan of folio 165r of an Arabic manuscript of the dīw\=an of Ibn San\=a' al-Mulk kept at 
the Bodleian Library in Oxford (shelf mark Marsh 236), showing the beginning of the poem on the transient of AD 1181/82.
The Arabic text on the supernova as presented in Arabic and English in Sect. 2.4 is seen here in lines 4 to 21 from
the top, i.e. one introductory line plus 17 verses with one line each.}
\end{figure}

\subsection{Commentary on the poem}
 
The poem itself uses the terms najm for “star”, al-Kaff al-Kha\d{d}īb for the constellation `The (Henna-)Dyed Hand' (in Cassiopeia), 
r\=a\d{s}id for “astronomer” and munajjim for “astrologer”, afl\=ak al-nuj\=um for “stars' sphere”, shih\=ab for meteor, 
and al-Sim\=ak al-R\=ami\d{h} for Arcturus. Hence, its author does have some astronomical knowledge.

For the astronomical interpretation, we go through the text line by line.

Introductory line: As already mentioned, the longer version of this sentence almost certainly does not 
come from the author of the poem or even the observer(s), 
but stems from a later compiler or copyist (Sect. 2.3). 
Anyway, in pre-modern times, any new transient celestial object could be considered a `comet' or,
if that technical term is not known or if its literal meaning was understood, a “star with a tail”. 
SN 1572 was likewise called a `comet' by various observers (see e.g. SG02). 
The compiler or copyist adds “the like of which had not been seen before”, which may point to an unusual transient, 
since true comets were relatively frequent (roughly one per decade) and known also among laymen.

V. 1 \& 2: Already here, the object is called najm, i.e. “star”: “the stars in the sky have increased in number”, 
which may well refer to a stationary point source (star) rather 
than an extended object (comet). The text in verse 2, “[The sky] adorned itself with the star [najm]”, 
may well point to an exceptional (bright?) object. Four Yemenite records 
on SNe 1006, 1572, and 1604 all use the term {\it najm} 
(Rada \& R. Neuh\"auser 2015, R. Neuh\"auser et al. 2016, R. Neuh\"auser et al. 2017a).
According to Kunitzsch (1995), the two terms {\it najm} and {\it kawkab} were used for stars in our sense,
{\it kawkab} also for planets and comet heads.
The term “najm” for “star” in verse 2 is used in singular, i.e. for just one new star (and not several stars or even meteors) 
among “the stars in the sky”.

V. 3: This play of words (“The Dyed Hand ... put on a (signet) ring”) clearly points 
to the constellation “The Henna-Dyed Hand”.
The Arabic constellation “Dyed Hand” is composed of the five bright stars 
in Cassiopeia: $\alpha, \beta, \gamma, \delta$, and probably $\zeta$ Cas (see next subsection for a detailed discussion of this position).
It again mentions indirectly that the object was a transient (“used to be unadorned” ... now “adorned”), 
possibly still present when the poem was written.
The new bright star is compared to a “(signet) ring”, possibly some gemstone (“adorned”) on the ring or 
something special or bright on one finger (signet or seal ring), 
even though a star would not have the form of a `ring'.
The phrase “adorned itself with a star ... it smiled through it” (v. 2) also indicates a positive interpretation by the poem,
which would be unusual for a cometary-like phenomenon.

V. 4: The “sky's hand” again points to “The Dyed Hand” (in Cas).
The phrases “adorned itself with a star” (v. 2) and “sky's hand boast of its star” 
(v. 4) might show that the new star was brighter than those five bright stars of Cassiopeia,
i.e. brighter than 2.25 mag ($\alpha$ Cas, the brightest) -- this is indeed bright
enough for serendipitous discovery (e.g. R. Neuh\"auser \& D.L. Neuh\"auser et al. 2024).

V. 5 \& 6: “this star has perplexed astronomers and astrologers alike ... They are in disagreement over it and 
stammering” -- this most certainly again points to the fact 
that the object seen was of very unusual nature, e.g. a transient without a tail and/or very bright 
(“perplexed astronomers”), so that the astrological interpretation was uncertain 
(e.g. no similar previous case). For the distinction between “astrologer” (munajjim) and “astronomer” 
(r\=a\d{s}id, literally “observer”, and other terms derived from the same root), see Rashed (2017, pp. 778-782).

V. 7 \& 8: 
"We can see how you buried a spear on the horizon, leaving its butt behind and throwing its sharp head."
The term “rum\d{h}” (“spear/lance”) used in verse 7 may already point the reader to 
al-Sim\=ak al-R\=ami\d{h} in verse 11, the Arabic name 
(“rum\d{h}” meaning “spear/lance”)
for both Arcturus ($\alpha$ Boo) alone and the small asterism with $\alpha$ and $\eta$ Boo.
(According to al-\d{S}\=ufī, Arcturus ($\alpha$ Boo)
together with the faint star $\eta$ Boo west of it, was considered a spear/lance.)
$^{c}$Abd al-\d{H}aqq (Ibn San\=a' al-Mulk 1958, p. 651) considers the 
butt and head of the spear 
to represent the star and its tail, which might indicate a comet.
However, only one of the two (“its sharp head”) ends up in the sky 
(the “butt” remains on the ground), so that this verse could also apply to a star without a tail. 
As seen from Cairo in Aug and Sep 1181 in the evenings, indeed, $\eta$ Boo is near the horizon, 
while $\alpha$ Boo (Arcturus) is visible (and sets soon).
Verse 8 reminds us that we should not read too much into the details of the poetic imagery of the text, including verse 7: 
“But this is just an error due to my fantasy, and a mistake due to my illusions.” 

V. 9: “The star is your father, who left his abode out of longing for you and in order to greet you.” 
The honorary title (laqab) of Saladin's father Ayy\=ub was Najm al-Dīn 
(literally “the star of religion”). 
Saladin's father Najm al-Dīn Ayy\=ub had died in AD 1173. 
Identifying a person after their death with a star is unusual in Arabic poetry, 
but given that the term najm for `star' was part of his name, referring to him in a poem dedicated to Saladin 
and dealing with a star was almost unavoidable for the poet.  

While allusions to meteors and Arcturus sometimes constitute topoi typical for panegyric poems, 
we can try to obtain additional, more speculative conclusions also from the remaining verses.

V. 10: 
“The stars' spheres” are the lunar and planetary spheres plus the (outer) stellar sphere (all supra-lunar).
The “meteors” (all meteorological phenomena), however, 
are located in the sub-lunar sphere (below the moon, in the Earth atmosphere).
Now, both “the stars' spheres” and “their meteors”, i.e. both the supra- and sub-lunar domains,
come to aid Saladin.
  
Transients (“meteors” and all other `meteorological', i.e. variable phenomena including what we today call comets) 
were often interpreted as portents for 
events on Earth, usually as negative signs in case of comets. 
But here, the transient object of AD 1181/82 is interpreted positively (e.g. vers 2) for Saladin. 
This can also indicate that the transient was unusual, i.e. neither a meteor nor a comet in today's sense. 

V. 11: “How often did Arcturus brandish its spear to strike your enemy until it was almost shattered to pieces!” 
The image of Arcturus (Arabic al-Sim\=ak al-R\=ami\d{h}) -- 
lit. “the lance-bearing sky-raiser” -- coming to the aid of the sovereign is quite common in Arabic poetry and should not be misunderstood 
as necessarily indicating the position or 
type (spear/lance, meteor?) of the phenomenon.
That the transient new object is compared here in a poetic way to Arcturus, and not any other star, 
may possibly indicate that the new star and Arcturus had a similar 
brightness (or less likely similar color). Arcturus has a visual brightness of V=0.16 mag (and a color index of B$-$V=1.14 mag). 

V. 13: “You rose until it was impossible to rise any further; you progressed until it was impossible to progress any further.”
The term “You” seems to point to Saladin, given the context.
More speculative, it may also be possible that the new star is meant (as well):
It may be a reflection of the light curve of the transient object, which has first risen to a maximum, 
then may have stayed around maximum for some time, but then needed to decline down to invisibility. 
The phrase “you rose until it was impossible to rise any further” would then mean that the poem was written after the 
brightness maximum during the decline phase. This is consistent with our dating of the composition of the poem (above), 
namely that Ibn San\=a' al-Mulk wrote it 
between December 1181 and May 1182.
Alternatively, the phrase “impossible to rise any further” can also just point to the star's high altitude.
The second part of the verse is a repetition of the first part.

V. 16: The second half of the verse alludes to Quran 16, 16: 
“And landmarks. And [men] can guide themselves by the stars” (translation by Alan Jones),
i.e. that stars can guide people for navigation purposes.

V. 17: Saladin is considered to be the “most elevated  ... in the sky”,
metaphorically brighter than all planets and stars including the new bright star.

At the latest after verse 17, there is no information anymore on any topics related to the transient of AD 1181.
The poet moved from the transient of AD 1181 to the father of Saladin, the spear/Arcturus, 
the spheres, and the lights of the sky and the stars (verses 12-16). 

All information from the poem itself (including its dating) is fully consistent with SN 1181: 
an unusual (relatively bright) object seen around AD 1181/82 in or near the five bright stars of Cassiopeia.

\subsection{Astronomical conclusions}

Here, we will consider object type, duration, brightness, color, and position of
the new star of AD 1181/82 according to the poem from Cairo.

{\bf Object type:}
The Arabic poem clearly speaks of a {\it star} by using the term {\it najm} several times,
and there is no evidence for motion or extension.
The four Arabic records from Yemen on SNe 1006, 1572, and 1604 all use the term {\it najm}, 
pointing to a stationary {\it stellar} transient (Rada \& R. Neuh\"auser 2015, R. Neuh\"auser et al. 2016, R. Neuh\"auser et al. 2017a).
Hence, we deal with the transient object of AD 1181/82 -- as also recorded in East-Asia -- as the subject of our poem.
This is usually named SN 1181.

{\bf Duration:} The observational time span in China from 1181 Aug 6 for a total of 185 days until 1182 Feb 6 is 
not inconsistent with the dating of our poem between December 1181 and May 1182 (the recitation before Saladin).
  
{\bf Brightness:} 
Verse 3 (`The Dyed Hand used to be unadorned, but when time became adorned because of you, it put on a (signet) ring.')
effectively says that the new star was brighter than the five bright stars of the constellation `Dyed Hand' in Cassiopeia,
i.e. brighter than its brightest star ($\alpha$ Cas with V=2.25 mag). \\
SG02 (p. 112) concluded that the star was very roughly around 0 mag or slightly brighter at maximum. 
This would be consistent with the comparison with Arcturus in our poem (V=0.16 mag), 
if that was really meant to indicate brightness, which is somewhat speculative. 
Since Arcturus is the 6th brightest star visible in Egypt, 
the new star at a similar brightness would also have been among the brightest stars (“astronomers and astrologers ... stammering”). 

When the new star became invisible, it was very likely around 5-6 mag, the limit for naked-eye detection.

Since the new star was mentioned in a poem completed in, after, or since December 1181
and recited before Saladin (and probably several to many others) in or before May 1182,
the new star must have been known to the guests present, so that it had an appreciable brightness
(if not during the recitation period, then at least some time ago).

{\bf Color:} The Japanese Azuma Kagami history reports directly in connection with the new star on 1181 Aug 7: 
“[like] Saturn-star color blue red, it had horned ray/s” 
(our literal translation). 
This was interpreted as to mean that the new star was similar as Saturn 
in brightness (e.g. SG02):
Saturn has $-0.5$ to 1.3 mag. 
A “blue-red” color could point to some colored rays seen from the bright new star due to scintillation.
(The phrase “horned ray/s” -- Classical Chinese does not distinguish between singular and plural --
can indicate scintillation and/or deformation due to low altitude and/or relatively large brightness.)
Whether the mentioning of Arcturus in our poem is a comparison of the new star in color with Arcturus 
might be too speculative.

Finally, let us now consider the transient of SN/nova 1181.

{\bf Position:} 
The remnant of SN/nova 1181 is not yet identified with full certainty: While SG02 preferred the SNR 3C58 near $\epsilon$ Cas, 
recently Oskinova et al. (2020), Ritter et al. (2021), Fesen et al. (2023), and Schaefer (2023) argued 
for the Parker star Pa-30 (IRAS 00500+6713, also in Cas, but not close to any bright star) 
as counterpart of SN 1181 (which might then be a SN Iax). Hence, positional information of the new star is most important.

The celestial location specified in the East-Asian records,
all reporting the far north near the circumpolar region (Xu et al. 2000, SG02),
may be consistent with a position within or near 
al-Kaff al-Kha\d{d}īb, 
`the dyed hand' or `the henna-colored/-dyed hand'.
Ibn Qutaybah (edition 1956, on Bedouin astronomy, 9th century) gave the following description for the constellation 
called {\it the Hand of Thurayy\=a'} (Thurayy\=a' means the Pleiades):
“this is the stretched hand of Thurayy\=a’,
five white [Arabic bī\d{d} for bright and/or white] stars in the Milky Way, 
opposite `the Fish' al-\d{H}\=ut” (Kunitzsch 1961, p. 72, in German).
Since the brightest of the stars to be considered here, $\alpha$ Cas, is not white
(B$-$V=1.18 mag, i.e. appearing orange-red), the Arabic term bī\d{d} here more likely
was meant for `bright' instead of `white'.
Sometimes, the term al-Kaff al-Kha\d{d}īb for `the dyed hand' was applied to just one of those five stars, 
namely $\beta$ Cas, probably because this is the only astrolabe star among the five (Kunitzsch 1961, p. 72, following al-\d{S}\=ufī 77, 23).

According to, e.g., W.B. Adams (2011, figure 1) and D.K. Adams (2021, p. 283), 
the Arabic and Bedouin constellation al-Kaff al-Kha\d{d}īb (“Dyed Hand”) 
would consist of $\alpha, \beta, \gamma, \delta$, and $\epsilon$ Cas (forming the letter M or W in Cassiopeia).
However, al-\d{S}\=ufī wrote that `the Arabs call the brightest of these stars (of Cassiopeia) the (henna-)dyed hand'
(al-\d{S}\=ufī 77, 17, cited after Kunitzsch 1961, p. 72).
Since Ibn Qutaybah speaks of {\it five} stars for al-Kaff al-Kha\d{d}īb, 
and al-\d{S}\=ufī of the {\it brightest},
we have to identify the five brightest stars in Cassiopeia for the constituents of al-Kaff al-Kha\d{d}īb --
i.e. the five stars taken to be the brightest in historical time.
According to Ptolemy's Almagest in the Greek (Toomer 1984) 
as well as the Arabic translation (both the al-\d{H}ajj\=aj and the Is\d{h}\=aq traditions,
see Kunitzsch 1986), and also according to al-\d{S}\=ufī himself (see Hafez 2010), 
the five stars considered to be and listed as the brightest were
$\alpha, \beta, \gamma, \delta$, and $\zeta$ Cas: 
The star $\gamma$ Cas with `brighter than (average) 3rd' mag (2.7 mag) was the (then) brightest in Cas
(today known as eruptive variable star with V=1.4-3.3 mag on AAVSO since 1936);
then $\alpha, \beta$ and $\delta$ are given as 3.0 mag; and 
$\zeta$ Cas was given as `brighter than (average) 4th mag' (ca. 3.7 mag).
These five brightest stars are followed in all the above catalogs by (at least)
two stars with 4th mag ($\eta$ and $\epsilon$ Cas);
further three stars with 4.0 or 4.3 mag, namely $\iota, \theta$, and $\kappa$ Cas.
Thus, a clear identification of the historically five brightest stars in Cas is given,
which are the {\it historical} constituents of al-Kaff al-Kha\d{d}īb: 
$\alpha, \beta, \gamma, \delta$, and $\zeta$ Cas.

The magnitudes and color indices of these stars according to modern measurements are as follows: \\
$\alpha$ Cas (Schadar) has V=2.25 mag (on AAVSO: V=1.5-3.1 mag) and B$-$V=1.18 mag (orange-reddish),  \\
$\beta$ Cas (Caph) has V=2.28 mag and B$-$V=0.38 mag (green, but seen whitish),  \\
$\gamma$ Cas has V=2.38 mag (eruptive variable star, on AAVSO: V=1.4-3.3 mag) and B$-$V=$-0.10$ mag (blue), \\
$\delta$ Cas (Ruchbah) has V=2.68 mag and B$-$V=0.16 mag (white), and \\
$\zeta$ Cas has V=3.68 mag and B$-$V=$-0.19$ mag (blue). \\
The star $\epsilon$ Cas (Segin) has V=3.35 mag and B$-$V=$-0.13$ mag (blue), i.e. it is the fifth brightest according to
modern measurements, but here, we have to consider which five stars were considered to be the brightest in historical time.

It may remain uncertain whether the observers in the past considered just the smallest possible polygon or ellipse 
around those five stars as area of the Dyed Hand,
or somewhat more including the neighborhood -- of course limited by (or half-way to?) neighboring constellations.
Ptolemy in his Almagest and, following suit, also al-\d{S}\=ufī listed 13 stars inside the main figure of Cassiopeia and none around it 
(called Dh\=at al-Kursī by al-\d{S}\=ufī, literally meaning “the one with the chair” referring to Cassiopeia sitting on a chair).
We display this area in Fig. 2.

\begin{figure*}
\centering
\includegraphics[angle=0,width=1.7\columnwidth]{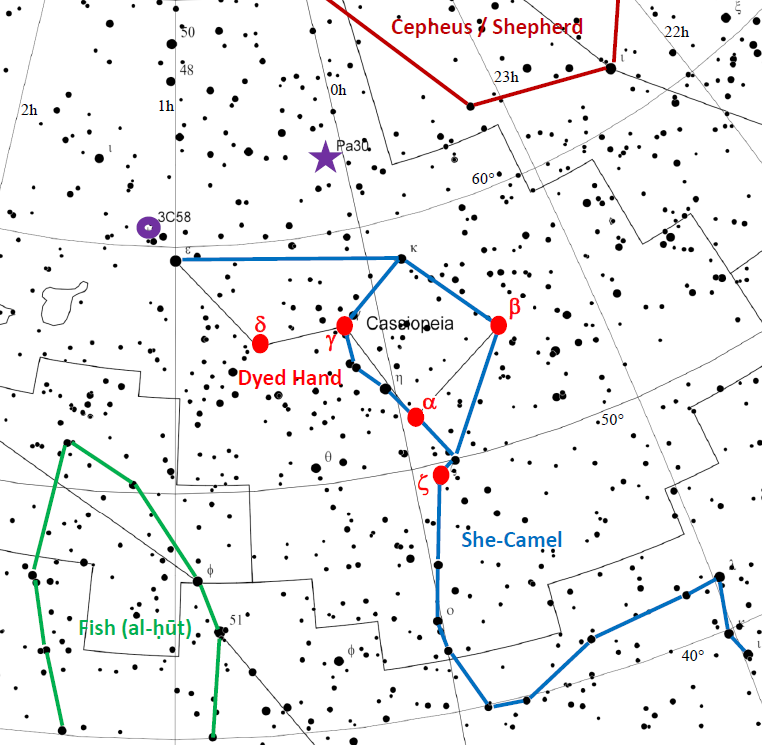}
\caption{{\bf Sky area Cassiopeia:} The relevant area on sky with al-Kaff al-Kha\d{d}īb 
(Dyed Hand: 5 stars in Cassiopeia: $\alpha, \beta, \gamma, \delta, \zeta$ Cas)
in red and the Bedouin constellation She-Camel in blue (Cas and And, we display only the bright, certain stars of this asterism), 
as well as other Arabic and/or Bedouin constellations nearby, namely Shepherd (Cepheus, top, north, al-r\=a$^{c}$ī)
and the other Fish (And/Per, bottom left, south-east, al-\d{h}\=ut). The two suggested remnants of the transient of AD 1181, 
3C58 and Pa-30, are marked in pink. 
We indicate the borders between the IAU constellations, their skeleton lines, equatorial coordinates, and names of a few bright stars
(as well as the five stars forming the letter M/W in Cas by black skeleton lines).
All stars down to ca. 6 mag are shown with sizes indicating their brightness. The figure was generated with Carte du Ciel.}
\end{figure*}

In sum, the area of $\alpha, \beta, \gamma, \delta$, and $\zeta$ Cas 
can be considered the Dyed Hand mentioned 
as position of the new star. This is by-and-large consistent with both the previously suggested remnants:
While 3C58 is close to $\epsilon$ Cas, the Parker star Pa-30 
is already midway 
between the Arabic constellations `Dyed Hand' (in Cas) and `Shepherd' (in Ceph), see Fig. 2.

\section{A new Arabic text on SN 1006 by al-Maqr\textit{\={\i}}z\textit{\={\i}}}
 
In an Egyptian chronicle by the prolific Cairene polyhistor 
A\d{h}mad ibn $^{c}$Alī al-Maqrīzī (AD 1364-1442),
we could identify a new record on SN 1006.
His three volumes history of the Fatimid dynasty (in Egypt since AD 969) 
is entitled “Admonition of the Pious -- Information concerning the Fatimid Imam-Caliphs” 
(Itti$^{c}$\=a\d{z} al-\d{h}unaf\=a' bi-akhb\=ar al-a'immah al-f\=a\d{t}imiyyīn al-khulaf\=a'). 
Under the year AH 396 (AD 1005 Oct 8 to 1006 Sept 26) 
we find the report on a new star in connection with
the insurrection of the Sunnite rebel Ab\=u Rakwah who rose in the Cyrenaika (Eastern Libya) 
against the Shiite (Ismaili) Egyptian caliph al-\d{H}\=akim. 
We first quote the report in Arabic and then give our English translation.

Arabic: wa-k\=ana fī \d{z}uh\=uri Abī Rakwatin \d{t}ala$^{c}$a kawkabun lah\=u dhu'\=abatun fa-k\=ana yu\d{d}ī'u ka l-qamari 
wa-lah\=u barīqun wa-lama$^{c}$\=anun wa-yaqw\=a wa-yakthuru n\=uruh\=u wa-amru Abī Rakwatin yashtaddu wa-ya$^{c}$\d{z}umu. 
fa-aq\=ama h\=adh\=a l-kawkabu shuh\=uran thumma \d{d}ma\d{h}alla n\=uruh\=u wa-\d{d}a$^{c}$ufa lama$^{c}$\=anuh\=u wa-akhadha amru 
Abī Rakwatin yanqu\d{s}u wa-ya\d{d}$^{c}$ufu il\=a an ukhidha asīran fa-gh\=aba l-kawkabu wa-lam yura ba$^{c}$da dh\=alika. 
fa-k\=ana sha'nu h\=adh\=a l-kawkabi fī dal\=alatihī $^{c}$al\=a Abī Rakwatin min a$^{c}$jabi l-$^{c}$ajabi.

English: “When Ab\=u Rakwah rose in revolt,
a star [kawkab] with a tail [dhu'\=abah] appeared. 
It shone like the moon with brightness and gleam and its light strengthened and increased so long 
as Ab\=u Rakwah's cause got on well and became ominous. This star [kawkab] remained (some) months; 
then its light dwindled and its gleam faded away while 
Ab\=u Rakwah's cause began to weaken and lost in strength until he was taken prisoner. 
Then the star [kawkab] disappeared and was no longer seen. 
The story of this star [kawkab] with its hint at Ab\=u Rakwah was one of the most astonishing things.”

Astronomical interpretation on light curve and brightness: 
The Arabic term `kawkab' was used for planets, stars, and comet heads (Kunitzsch 1995),
similar in Syriac `kawkb\=o' (D.L. Neuh\"auser et al. 2021);
Ibn Sīn\=a from Persia also used it for SN 1006: `kawkab min al-kaw\=akib', i.e. `a star among the stars' (R. Neuh\"auser et al. 2017b).

It is mentioned explicitly that “its light strengthened and increased so long as Ab\=u Rakwah's cause got on well”. 
This may indicate that the increase in light was indeed observed after the first detection. 
The phase “It shone like the moon with brightness” would indicate a peak brightness like the moon, 
but it could also mean that it did produce shadows like the moon -- 
$^{c}$Alī ibn Ri\d{d}w\=an compared the brightness with the “quarter moon”. 
That the object is called “a star with a tail (dhu'\=abah)” should not worry us -- 
this can be due to strong scintillation and deformation at large brightness.

According to al-Maqrīzī, 
Ab\=u Rakwah's troops threatened Alexandria and gained a victory over an Egyptian army at the end of May AD 1005 (AH 395 Sha$^{c}$b\=an 15). 
After his defeat on AD 1006 Aug 30 the rebel was taken prisoner in a Nubian monastery where he had fled 
(AD 1006 Dec 25 to 1007 Jan 22/AH 397 Rabī$^{c}$ I). 
These dates correspond with the report on SN 1006 by the Christian author Ya\d{h}y\=a ibn Sa$^{c}$īd al-An\d{t}\=akī (d. AD 1065/AH 458), 
for the first time analyzed by Cook (1999); see also R. Neuh\"auser et al. (2017a) and Halm (2003, pp. 210-214). 
According to al-An\d{t}\=akī the star was seen four months from AD 1006 May 2 (AH 396 Sha$^{c}$b\=an 2) on. 
Several Arabic reports agree that the new star was discovered in the first days of May 1006 and then seen for three to six months (SG02). 
Two reports from Yemen explicitly give a significantly earlier date (mid-Rajab AH 396) corresponding to 1006 April $17 \pm 2$;
also the record from St. Gallen is consistent with a first detection in April 1006 (R. Neuh\"auser et al. 2017a).
Al-Maqrīzī's record may be consistent with a visibility of SN 1006 until Sep 1006.

Al-Maqrīzī was not a contemporary of these events, but his report is based upon earlier sources. 
The Fatimid dynasty whose history he writes had been a heterodox Shiite (Ismaili) regime; 
that is the reason why after its fall and the Sunnite restauration 
under sultan Saladin (since AD 1169) it fell victim to a kind of damnatio memoriae, 
as a result of which nearly the complete literature of the Fatimid 
epoch was destroyed or fell into oblivion.

Fortunately al-Maqrīzī still had first-hand information at his disposal: 
he amply quotes (or at least paraphrases) the “Chronicle of Egypt“ (Akhb\=ar Mi\d{s}r) 
by Mu\d{h}ammad ibn $^{c}$Ubaydall\=ah al-Musabbi\d{h}ī (AD 977-1029/AH 366-420), a courtier of the Fatimid caliph al-\d{H}\=akim (r. AD 996-1021). 
Of al-Musabbi\d{h}ī's work only a small fragment, dealing with the years AD 1023-25, has survived in a manuscript in the Escorial library, 
but in al-Maqrīzī's work large parts of it are preserved. This makes our text 
another testimony of a contemporary witness of SN 1006 from Egypt 
beside the above mentioned al-An\d{t}\=akī (d. AD 1065), 
who left Cairo in AD 1014/1015 (AH 405) for Syria and 
$^{c}$Alī ibn Ri\d{d}w\=an (ca. AD 988-1061), who both observed SN 1006, when they were young. 

\section{Summary}

We present historical transmissions in Arabic language from Cairo, Egypt, for both SN/nova 1181 and SN 1006. 
The poem by Ibn San\=a' al-Mulk from Egypt is the first text from Arabia for SN/nova 1181, 
written in between December 1181 and May 1181, as re-dated by us (Sect. 2.3),
in order to praise the famous sultan Saladin.

The Arabic poem presented here yields valuable information on the transient of AD 1181/82, 
in particular on its nature (a stellar object, Arabic: najm), dating (Dec 1181 to May 1182), 
celestial position (in Cassiopeia), 
and brighter than the brightest star in Cas ($\alpha$ Cas with V=2.25 mag) --
and maybe also like Arcturus. 
Also, the very fact that it was indeed seen by laymen (the poet)
and discussed intensively among astronomers and astrologers 
in Egypt indicates a relatively 
strong brightness and/or lengthy visibility period.
Several of the SN criteria by SG02 (see Sect. 1.1) are fulfilled by the Arabic poem alone,
in particular low Galactic latitude (in Cassiopeia),
and very likely also stationarity as a true star without angular extension (Arabic: najm) --
most importantly, it is an additional record independent of those from China and Japan.

Even though of uncertainties involved in the interpretation of poems,
here, the recorded realistic facts, the technical terms used, the discussions among experts, the recitation before Saladin,
and the fact that the author was probably an eyewitness of the unusual transient,
all show that we deal with a real new star.
We expect that more texts from Cairo or elsewhere in Arabia and Persia on this new star were written
around that time, maybe with more details on dating, brightness, or even position. 

In addition, we present another report from Arabia on SN 1006, for which several other Arabic reports were found before. 
The Egyptian chronicler al-Maqrīzī (AD 1364-1442) from Cairo described the new star of AD 1006
quite certainly by paraphrasing the “Chronicle of Egypt” by al-Musabbi\d{h}ī (AD 977-1029/AH 366-420),
who could have been another Egyptian eyewitness of SN 1006. This report does not provide any new 
astronomical information on the supernova, but confirms previously known aspects like the brightness somewhat comparable to the moon, 
increase and decrease in brightness during AD 1006, and a total duration of several months.

We should not be surprised that in both texts, the object was called a “comet” (“star with tail”). 
In the case of SN/nova 1181, this term was not used by the contemporaneous author of 
the poem, an eyewitness of SN/nova 1181, but by a later editor of only one recension, who added a headline (Sect. 2.3);
the compiler thought that was reported as a “new star” was a star with a tail
-- he had neither known nor seen any tail-less bright new star.
And in the case of SN 1006,
only the introductory sentence has “a star with a tail” (maybe deformation at strong brightness as reported elsewhere),
while it was then several times called `kawkab' for `star'.

\smallskip

{\bf Acknowledgements.}
RN and HH would like to thank Klaus Werner (U T\"ubingen) for putting us into contact on the new Arabic report of SN 1006. 
JGF would like to thank the staff of the Bodleian Library for providing scans of some of their Arabic manuscripts, 
including the image shown in Fig. 1, as well as his colleagues at the University of M\"unster for 
checking his edition and translation of the poem on SN 1181. \\
Author contribution: JGF noticed that the celestial transient of AD 1181 is described in our poem,
he edited, translated, and re-dated the poem (Sects. 2.1-2.4); 
HH found the text on SN 1006 and wrote most of Sect. 3;
DLN contributed in particular to the astronomical interpretation of the new texts (Sects. 2.5, 2.6, 3);
RN contributed Sect. 1 and Fig. 2; Sects. 2.5, 2.6, and 4 were written together;
all authors read and commented on the whole paper.

{}

\end{document}